\documentclass[11pt]{article}

\usepackage{SIGACTNews}

\newcommand{\COLUMNsplit}{\vskip 2em\hrule \vskip 3em}

\newcommand{\COLUMNguest}[2]{%
  \begin{center}%
   {\LARGE\bf #1 \par}%
    \vskip 1.5em%
    {\Large
      \lineskip .75em%
      \begin{tabular}[t]{c}%
        #2
      \end{tabular}\par}%
    \vskip 1.5em%
  \end{center}%
  \par}

\usepackage[german, english]{babel}
\usepackage[all]{xy}
\usepackage{bbm}
\usepackage{latexsym}
\usepackage{amsmath}
\usepackage{amssymb}
\usepackage{amsthm}
\usepackage[dvips]{graphicx}

\newcommand{\npskip}{\vspace{-\parskip}}

\renewcommand{\phi}{\varphi} 

\newcommand{\imp}{\ensuremath{\rightarrow}}
\newcommand{\biimp}{\ensuremath{\leftrightarrow}}
\newcommand{\et}{\ensuremath{\wedge}} %
\newcommand{\ou}{\ensuremath{\vee}} %
\newcommand{\notmodels}{\ensuremath{\mathrel{\mathpalette\notmodelss\models}}}
  \newcommand{\notmodelss}[2]{\ooalign{$\hfil#1\mkern1mu/\hfil$\crcr$#1#2$}}
\newcommand{\oif}{\ensuremath{\;\Rightarrow\;}} %
\newcommand{\si}{\ensuremath{\;\Leftarrow\;}} %
\newcommand{\ssi}{\ensuremath{\;\Leftrightarrow\;}} %

\newcommand{\sem}[1]{[\![#1]\!]}

\newcommand{\too}{\longrightarrow}
\newcommand{\inv}{{^{-1}}}

\newcommand{\bbtwo}{\mathbbm{2}}

\newcommand{\acal}{\mathcal{ A}} 
 
\newcommand{\ccal}{\mathcal{ C}} 
\newcommand{\dcal}{\mathcal{ D}}

\newcommand{\ocal}{\mathcal{ O}} 
\newcommand{\pcal}{\mathcal{ P}}

\newcommand{\xcal}{\mathcal{ X}}

\newcommand{\Asf}{\mathsf{A}}
\newcommand{\Bsf}{\mathsf{B}}

\newcommand{\Xsf}{\mathsf{X}}

\newcommand{\Poset}{\mathsf{Poset}}

\newtheorem{theorem}{Theorem}[section]
\newtheorem{theorem-conj}{Theorem (Conjecture)}[section]

\newtheorem{lemma-conj}[theorem]{Lemma (Conjecture)}

\theoremstyle{definition}
\newtheorem{definition}[theorem]{Definition}

\theoremstyle{remark}

\newcommand{\id}{\mathrm{id}}
\newcommand{\iso}{\cong}
\newcommand{\op}{{^{\mathrm{op}}}}

\newcommand{\Set}{\mathsf{Set}}

\newcommand{\Sob}{\mathsf{Sob}}

\newcommand{\Frm}{\mathsf{Frm}}
\newcommand{\BA}{\mathsf{BA}}
\newcommand{\CABA}{\mathsf{CABA}}
\newcommand{\DL}{\mathsf{DL}}
\newcommand{\CDL}{\mathsf{CDL}}
\newcommand{\Stone}{\mathsf{Stone}}
\newcommand{\Spec}{\mathsf{Spec}}

\newcommand{\Alg}{\mathsf{Alg}}

\newcommand{\Coalg}{\mathsf{Coalg}} %

\theoremstyle{remark}

\theoremstyle{definition}
\newtheorem{defn}[theorem]{Definition}

\renewcommand{\rho}{\varrho}

\newcommand{\sBox}{\mbox{$\Box\hspace{-1.45ex}\Box$}}

\newcommand{\fca}{{\tilde P}} %
\newcommand{\fac}{{\tilde S}} %

\renewcommand{\Asf}{A}
\renewcommand{\Bsf}{B}
\renewcommand{\Xsf}{X}
\renewcommand{\bbtwo}{2}

\title{SIGACT News Logic Column 15}
\author{Riccardo Pucella\\
Northeastern University\\
Boston, MA 02115 USA\\
riccardo@ccs.neu.edu}
\date{}

\begin{document}

\SIGACTmaketitle

Some comments about the last Logic Column, on nominal logic. 
Pierre Lescanne points out that the terminology ``de Bruijn levels''
was introduced in the paper \emph{Explicit Substitutions with de
  Bruijn's Levels}, by Pierre Lescanne and Jocelyne Rouyer-Degli,
presented at the 1995 RTA conference. 
He also points out that Stoy diagrams were probably invented by Stoy,
but appear in work by Bourbaki as early as 1939 (published in 1954).
Merci, Pierre.

That article also initiated what is bound to be an interesting
discussion. 
The critique of higher-order abstract syntax in that article prompted
Karl Crary and Robert Harper to prepare a response to the leveled
criticisms. 
The response should appear in an upcoming Column.

In this issue, Alexander Kurz describes recent work on the topic of
specifying properties of transition systems. It turns out that by
giving a suitably abstract description of transition systems as
coalgebras, we can derive logics for capturing properties of these
transition systems in a rather elegant way. I will let you read the
details below.

I am always looking for contributions. If you have any suggestion
concerning the content of the Logic Column, or if you would like to
contribute by writing a column yourself, feel free to get in touch
with me. 

\COLUMNsplit

\COLUMNguest{Coalgebras and Their Logics\footnote{\copyright{} Alexander Kurz, 2006.}}
   {Alexander Kurz\\ University of Leicester, UK}

\section{Introduction}

Transition systems pervade much of computer science. This article
outlines the beginnings of a general theory of specification languages
for transition systems. More specifically, transition systems are
generalised to coalgebras. %
Specification languages together with their proof systems, in the
following called (logical or modal) calculi, are presented by the
associated classes of algebras (e.g., classical propositional logic by
Boolean algebras). Stone duality
will be used to relate the logics and their coalgebraic semantics.
The relationship between these notions can be summarised as
\begin{equation}\label{picture}
\xymatrix{\textrm{systems} \ar@{-}[r] & \textrm{coalgebras}\\
  \textrm{logic} \ar@{..}[u] \ar@{-}[r] & \textrm{algebras}
  \ar@{-}[u]_{\textrm{Stone duality}} }
\end{equation}

\medskip\noindent Let us have a closer look at the role of Stone
duality, which relates the class $\acal$ of algebras of a
propositional logic to the class $\xcal$ representing the carriers of
the transition systems (i.e., coalgebras). The duality is provided by two
operations $P$ and $S$
\begin{equation}
\label{eins}
\xymatrix{
  {\ \xcal} \ar@/^/[r]^{P} & {\acal \ } \ar@/^/[l]^{S} ,
   }
\end{equation}
where $P$ maps a carrier $X$ to its propositional theory and $S$ maps
an algebra $A$ to its `canonical model' ($PX$ and $SA$ are also
called the \emph{dual} of $X$ and $A$, respectively; the reason
for the terminology will become clear later).  The situation in
Diagram~\ref{eins} describes a perfect match of logic and semantics if
both models and theories can be reconstructed (up to isomorphism) from
their dual, formally, if $X\iso SPX$ and $A\iso PSA$ for all
$X\in\xcal$ and $A\in\acal$.

\medskip\noindent As we will explain below, the type of a class of
coalgebras is an operation $T:\xcal\to\xcal$.  This suggests that, in
the same way as a logic for $\xcal$ is given by the algebras in
$\acal$, a logic for $T$-coalgebras is given by the algebras for the
corresponding operation $L$ on $\acal$.
The Stone duality~(\ref{eins}) then lifts to $L$-algebras and
$T$-coalgebras
\begin{equation}\label{drei}
\xymatrix{
  {\Coalg(T)} \ar@/^/[r]^{\fca} \ar[d]
  & {\Alg(L)\ } \ar@/^/[l]^{\fac} \ar[d]\\
  {\ \cal X} \ar@(dl,ul)[]^{T} \ar@/^/[r]^{P}
   & {\acal\ } \ar@/^/[l]^{S} \ar@(dr,ur)[]_{L} 
  \\
}
\end{equation}
The logical interpretation of the upper duality is that the logic
corresponding to $\Alg(L)$ is (strongly) complete and characterises
bisimilarity wrt the transition system $\Coalg(T)$.
This will be explained in Section~\ref{sec:lc}.
Section~\ref{sec:sd} discusses one example of the dualities in
Diagram~\ref{eins} in detail and briefly sketches the more general
picture. The remainder of the introduction is devoted to a more
detailed exposition of the ideas above. Section~\ref{sec:outlook}
summarises what can be gained from this approach and outlines some
research directions. Definitions of the few notions from category
theory we need are collected in an appendix.

\subsection{Systems as Coalgebras} 
In its simplest form, a transition system
consists of a set $X$ and a relation $R\subseteq X\times X$. Denoting
by $\pcal$ the operation mapping a set to its powerset, a transition
system can equivalently be described by a map
\[
X\stackrel{\xi}{\too} \pcal X
\]
where $\xi(x)=\{y | xRy\}$ is the set of successors of $x$. 
The structure $(X,\xi)$ is a $\pcal$-\emph{coalgebra}.

\medskip\noindent This description of transition systems is very
flexible. Table~\ref{table:coalgebras} gives some examples of
$T$-coalgebras $(X,\xi)$ for suitable mappings $T:\Set\to\Set$. The
set $C$ appearing in some functors is a constant parameter denoting an
input or output alphabet.
\begin{table}
\renewcommand{\arraystretch}{1.4}
\begin{center}
\begin{tabular}{|l|l|l|}
  \hline
   & $T X$ & $X\stackrel{\xi}{\too} TX$\\
  \hline
  \hline
  1. & $C\times X$ & streams over $C$\\
  \hline
  2. & $C\times X + 1$ & finite or infinite lists over $C$\\
  \hline
  3. & $2\times X^C$ & deterministic automaton with input alphabet $C$\\
  \hline
  4. & $\pcal(C\times X)\iso(\pcal X)^C$ & $C$-labelled transition system\\
  \hline
  5. & $(1+ \dcal X)^C $ & probabilistic transition systems \\
  \hline
  6. & $2^{2^X}$ & predicate transformer\\
  \hline
\end{tabular}
\end{center}
\caption{Examples of Coalgebras}\label{table:coalgebras}
\end{table}
The models we have in mind are often systems with a distinguished
initial state (which we also call \emph{pointed} coalgebras or
\emph{processes}).  (1) A coalgebra $ X\to C\times X$ with specified
initial state $x_0$ is a process outputting the infinite stream
$(\mathit{head}(x_0),\mathit{head}(\mathit{tail}(x_0)),\ldots)$ of
elements of $C$, where $\mathit{head}:X\to C$ and $\mathit{tail}:X\to
X$ refer to the two components of $ X\to C\times X$. (2) Here 1
denotes a one-element set, which allows a process to stop; hence, in
addition to streams, one now also allows behaviours given by finite
lists.  (3) The 2 in $2\times X^C$ denotes a two-element set and the
$X\to 2$ part of the coalgebra expresses whether a state is accepting
or not; $X^C$ denotes the set of functions from $C$ to $X$ and $X\to
X^C$ calculates the successor state from a current state and an input
from $C$.  (4) Comparing with (1) and (3), $\pcal(C\times X)$ suggests
to think of the labels in $C$ as outputs and $(\pcal X)^C$ of the
labels as inputs, but both are isomorphic. (5) The
distribution functor $\dcal X$ maps $X$ to the set of discrete
probability distributions.
  (6) Coalgebras $X\to2^{2^X}$ are (in bijective correspondence to)
  predicate transformers $2^X\to 2^X$.

\paragraph{Bisimilarity } The crucial observation that makes the
coalgebraic point of view useful is that all of these type
constructors $T$ are functors and that, therefore, $T$-coalgebras come
equipped with a canonical notion of behavioural equivalence or
bisimilarity.

\medskip\noindent Let us explain this important point in more detail.
To say that an operation $T:\Set\to\Set$ is a functor means that $T$
is not only defined on sets but also on functions, mapping $f:X\to Y$
to $Tf:TX\to TY$.  Moreover, $T$ is required to preserve identities
$\id_X:X\to X$, and composition, $T(f\circ g)=Tf\circ Tg$. This allows
us to define a morphism of coalgebras $(X,\xi)\to(Y,\nu)$ as a map
$f:X\to Y$ such that $Tf\circ\xi = \nu\circ f$:
\begin{equation}\label{def:morphism}
\xymatrix{
X \ar[r]^{\xi}\ar[d]_{f} & TX \ar[d]^{Tf}\\
Y \ar[r]^{\nu} & TY
}
\end{equation}
Coalgebraic \emph{bisimilarity}\label{def-bisim}, or \emph{behavioural
  equivalence}, denoted $\simeq$, is now the smallest equivalence
relation that is invariant under all morphisms: Define $\simeq$ to be
the smallest equivalence relation containing all pairs $x\simeq f(x)$
where $f$ ranges over coalgebra morphisms and $x$ over the domain of
$f$.\footnote{This definition is equivalent to the following one. Two
  states $x,y$ in two coalgebras are bisimilar iff there are two
  coalgebra morphisms $f,g$ such that $f(x)=g(y)$.}

\medskip\noindent Coming back to our introductory example, $\pcal$ is
a functor if we let $\pcal f:\pcal X\to \pcal Y$ map a subset of $X$
to its direct image under $f$. It is now an instructive exercise to
show that coalgebraic bisimilarity agrees with the standard notion of
bisimilarity for (unlabelled) transition systems. For this, one shows
that a map is a coalgebra morphism iff its graph is a bisimulation
between the two transition systems.

\medskip\noindent In examples of Table~\ref{table:coalgebras} we have
the following.  (1,2) Two processes are bisimilar iff their outputs
are the same. (3) Two states are bisimilar iff they accept the same
language. (4) Here coalgebraic bisimilarity is the one known from
modal logic or process algebra and similarly for (5) and (6).

\paragraph{Final Coalgebras } It is often possible to characterise
bisimilarity by a single coalgebra, the so-called final coalgebra.
Formally, a coalgebra is \emph{final} if from any other coalgebra
there is a unique morphism into the final coalgebras. It follows that
two states in two coalgebras are bisimilar iff they are identified by
the unique morphisms into the final coalgebra.  In other words, the
final coalgebra, if it exists, provides a canonical representative for
each class of bisimilar states.

\medskip\noindent In many cases, apart from characterising
bisimilarity, final coalgebras are interesting objects in their own
right. Recognising these structures as final coalgebras allows to
reason about them, sometimes to great advantage, using coinduction
instead of induction. We will not pursue this issue any further here
but only mention some examples. In Table~\ref{table:coalgebras} the
final coalgebra is in (1) the coalgebra of streams over $C$, in (2)
the coalgebra of finite and infinite lists over $C$, in (3) the
coalgebra of all languages (with successors given by language
derivative). In our leading example of $\pcal$-coalgebras the final
coalgebra is the universe of non-well founded sets.

\subsection{Modal Logic} 
Let us now turn to logics for coalgebras. One would want such logics
to respect bisimilarity, that is, formulae should not distinguish
bisimilar states. We call such logics modal because it can be argued
that invariance under bisimilarity is the main feature of modal logic.
For example, a theorem of van Benthem states that modal logic is
precisely the fragment of first order logic that is invariant under
bisimilarity. Moreover, either by strengthening the logic allowing for
infinite conjunctions or be restricting the semantics to finitely
branching transition systems, modal logic characterises bisimilarity
in the sense that for each two non-bisimilar states there is a formula
distinguishing them. 

\medskip\noindent Let us first look at the usual modal logic for
unlabelled transition systems, ie,  $\pcal$-coalgebras.  It consists of
classical propositional logic extended by one unary operator $\Box$.
The interpretation of $\Box$ is that of a restricted universal
quantifier, more precisely, $\Box\phi$ holds in state $x$ iff $\phi$
holds in all successors of $x$. We write this as
\begin{equation}\label{sem-box-0}
  \sem{\Box \phi}_{(X,R)} = \{ x \mid xRy \oif
  y\in\sem{\phi}_{(X,R)} \} 
\end{equation}
  Having seen a logic for $\pcal$-coalgebras, can we generalise this
  to arbitrary functors $T$?

  \medskip\noindent Note first that, semantically, a modal operator
  transforms predicates into predicates. So we could say that a modal
  operator is a suitable operation
\[2^X\to 2^X\]
where $2^X$ denotes again the set of functions $X\to 2$, or
equivalently, the set of subsets of $X$. 
But we also want to capture that $\Box$ says something about the
immediate successors of a state, that is, about a single transition
step.  We therefore identify (one-step) modalities $\Box$ for $T$ with
so-called \emph{predicate liftings}\footnote{`Predicate lifting' because
  $\sBox$ lifts a predicate on $X$ to a predicate on $TX$.}
\begin{equation}\label{pred-lift}
  \sBox_X: 2^X\to 2^{TX}
\end{equation}
which give the semantics of a modal operator $\Box$ wrt a coalgebra
$(X,\xi)$ via
\begin{equation}
\xymatrix{
2^X & \ar[l]_{\xi^{-1}} 2^{TX} & 2^X \ar[l]_{\ \sBox_X}\\
}
\end{equation}
that is, dropping the subscripts,
\begin{equation}\label{sem-box}
  \sem{\Box\phi} = \xi^{-1}\circ \sBox{\sem{\phi}}.
\end{equation}
To recover (\ref{sem-box-0}) as a special case of (\ref{sem-box}), one
defines the corresponding predicate lifting as $\sBox Y = \{
Z\subseteq X \mid Z\subseteq Y\}$ for $Y\in 2^X$.

\medskip\noindent How do we guarantee that modal logics for coalgebras
given by predicate liftings are invariant under bisimilarity?  Simply
by requiring that predicate liftings $\sBox_X: 2^X\to 2^{TX}$
are natural transformations. Spelling out the definition of a natural
transformation this means that in the following diagram the right-hand
square commutes (for all $f:X\to Y$)
\begin{equation}\label{equ:nat}
  \xymatrix@C=60pt{
    2^X & \ar[l]_{\xi^{-1}} 2^{TX} & 2^X \ar[l]_{\ \ \sBox_X} \\
    2^Y \ar[u]_{f\inv} & 
      \ar[l]_{\nu^{-1}} 2^{TY} \ar[u]_{(Tf)\inv} & 
      2^Y \ar[l]_{\ \ \sBox_Y} \ar[u]_{{f}\inv}
  }
\end{equation}
If, moreover, $f:(X,\xi)\to (Y,\nu)$ is a coalgebra morphism, then
also the left-hand square commutes and, therefore, the outer rectangle
as well.  The proof of invariance of the logic under bisimilarity
(p.~\pageref{def-bisim}) is now a routine induction on the structure
of the formulae, where the case of modal operators $\Box$ is taken
care of by Diagram (\ref{equ:nat}).

\subsection{Logics as Algebras}
After having explained the basic notions of coalgebras and their
logics, in the remainder of the article, I will sketch a deeper
analysis of the situation.
It is based on the insight that logics for coalgebras are in fact
algebras and, moreover, that a logic perfectly captures the
coalgebras if the algebras and coalgebras are related by Stone
duality.

\medskip\noindent Observe that the modal logics discussed above come
in two stages. First, for any coalgebra $(X,\xi)$ we have the Boolean
algebra $2^X$, which corresponds to propositional logic. This logic is
then extended by modal operators $2^X\to 2^X$. Traditionally, these
algebras, called Boolean algebras with operators or modal algebras,
are thought of as given by a carrier $A$ plus boolean operators
$\bot,\neg,\et,\ou$ plus (possibly more than one) modal operator
$\Box$.
For example, the modal logic for $\pcal$-coalgebras can be given by
one unary modal operator $\Box$ that preserves top (true) and
conjunction
\begin{equation}\label{K}
\Box\top=\top \quad \quad \quad \quad \Box(a\et b) = \Box a \et \Box b \end{equation}
This example shows clearly the relationship between algebras and modal
calculi. On the one hand, (\ref{K}) is the equational definition of
the class of modal algebras. On the other hand, (\ref{K}) plus
equational logic provides a calculus for modal logic.  Since modal
logics are more commonly given by Hilbert calculi, we indicate briefly
that both calculi are equivalent in a rather straightforward way.

\medskip\noindent The usual Hilbert calculus of the modal logic for
$\pcal$-coalgebras, denoted \textbf{K}, has as axioms all
propositional tautologies and $\Box(p\imp q)\imp(\Box p\imp \Box q)$;
rules are modus ponens, substitution, and necessitation `from $\phi$
derive $\Box\phi$'. To compare \textbf{K} with the equational calculus
given by (\ref{K}), we write $\vdash_{\textbf{K}}\phi$ and
$\vdash_{\textsf{EL}}\phi=\psi$ for formulae derivable in \textbf{K}
and equations derivable in equational logic. One then shows that
$\vdash_{\textbf{K}}\phi \ssi {\vdash_{\textsf{EL}}\phi=\top}$ and
$\vdash_{\textsf{EL}}\phi=\psi \ssi
{\vdash_{\textbf{K}}\phi\biimp\psi}$. For example, the necessitation
rule is simulated on the equational side by the congruence rule of
equational logic and the first of the axioms~(\ref{K}).

\medskip\noindent Up to now we have only seen standard material from
modal logic. It will now be shown that modal algebras are algebras for
a functor. Since algebras for a functor are, in a precise sense, dual
to coalgebras this will allow us to relate modal algebras (and hence
modal calculi) in a uniform way to their coalgebraic semantics.

\medskip\noindent We start by observing that the two-stage process of
building a modal algebra can be made more explicit by saying that a
modal algebra is a Boolean algebra $\Asf$ with a finite-meet
preserving map $\Box: \Asf\rightharpoonup \Asf$. From a technical
point of view, it is inconvenient that $\Asf$ is a Boolean algebra but
$\Box$ is only a meet-semi-lattice morphism, which does not preserve
all of the Boolean structure. This is easily rectified: Modal algebras
are in one-to-one correspondence to algebras for the functor $L$ where
\begin{equation}\label{equ:def-L}
\textrm{$L\Asf$ is  the free Boolean algebra over $\Asf$
considered as a meet-semi-lattice.}
\end{equation}
This means that $L\Asf$ is determined by the property that for each
finite-meet preserving function $\Asf\rightharpoonup\Bsf$ there is a
unique Boolean algebra morphism $L\Asf\to\Bsf$ such that
\begin{equation}
\xymatrix{
L\Asf \ar[r] & \Bsf \\
\Asf \ar@^{->}[u] \ar@^{->}[ur] &
}
\end{equation}
commutes. It follows that Boolean algebra morphisms $L\Asf\to\Asf$
are in one-to-one correspondence with semi-lattice morphisms
$\Asf\rightharpoonup\Asf$. 

\medskip\noindent $L\Asf\to\Asf$ is an \emph{algebra for the functor}
$L$. This notion of an algebra for a functor dualises the notion of a
coalgebra, the arrows going in opposite directions: into the carrier
for algebras and out of the carrier for coalgebras (the appendix gives
a more formal statement of this duality).

\medskip\noindent Let us summarise the relationship between logics and
algebras in our example. Boolean algebras correspond to classical
propositional logic. A functor $L$ specifies an extension of
propositional logic with modal operators. The algebras for this modal
logic are the algebras for the functor $L$.
\renewcommand{\arraystretch}{1.4}\label{table:L}
\begin{center}
\begin{tabular}[h]{|l|l|}
  \hline
  class $\BA$ of Boolean algebras & classical propositional logic\\
  \hline
  functor $L:\BA\to\BA$ & modal operators + axioms \\
  \hline
  class $\Alg(L)$ of modal algebras & modal logic\\
  \hline
\end{tabular}
\end{center}

\subsection{Relating Algebras and Coalgebras via Stone Duality}

We have explained so far the two horizontal lines of the
picture~(\ref{picture}), namely systems as coalgebras and logics as
algebras. We now come to Stone duality, relating the two. 

\medskip\noindent We start by remarking that finitary modal logic does
not describe $\pcal$-coalgebras perfectly in the following sense.
First, finitary logics cannot characterise bisimilarity, that is, they
are not strong enough to distinguish all non-bisimilar states.
Furthermore, there are consistent modal logics that are incomplete in
the sense that there are no $\pcal$-coalgebras satisfying
them.\footnote{This phenomenon appears if the proposition letters of
  the modal axioms are interpreted as ranging over all subsets of the
  carrier of the model (Kripke frame semantics). It does not happen if
  proposition letters receive a fixed interpretation (Kripke model
  semantics).}  There are two ways to rectify this mismatch.

\medskip\noindent The first is based on the observation that $2^X$ is
not only a Boolean algebra but also has infinitary intersections, or
algebraically speaking, $2^X$ is a complete atomic Boolean
algebra.
This suggests that a perfect description of transition system requires
infinitary propositional logic. This is well-known in process algebra:
For infinitely branching transition systems Hennessy-Milner logic only
characterises bisimilarity if one allows infinite conjunctions.

\medskip\noindent Alternatively, instead of strengthening the logic by
infinitary constructs, one can modify the semantics to take the weaker
expressivity of the logic into account: One equips transition systems
with a notion of `admissible' or `observable' predicate.  For this,
one usually lets carriers consist not of sets but topological spaces
$(X,\ocal X)$.  The topology $\ocal X$ is a subset of $2^X$ encoding
which predicates on $X$ can be expressed by the logic.\footnote{For
  example, consider the modal logic \textbf{K} and a $\pcal$-coalgebra
  that is a tree with initial state $x_0$ having branches of any
  bounded length and one infinite branch. The subset of states
  reachable in bounded branches is not admissible. This corresponds to
  the fact that having an infinite branch is not expressible in the
  finitary logic \textbf{K}.}

\medskip\noindent In the first case, the algebras in
Diagram~\ref{eins} are complete atomic Boolean algebras, so
Diagram~\ref{eins} becomes
\begin{equation*}
\xymatrix{
  {\ \Set} \ar@/^/[r]^{P} & {\CABA \ } \ar@/^/[l]^{S} ,
   }
\end{equation*}
where $PX=2^X$ and $S$ maps an algebra to its set of
atoms.\footnote{$a$ is an atom if $\bot<a$ and $\bot<b\le a\oif b=a$.}

\medskip\noindent In the second case, the algebras are Boolean
algebras. The corresponding spaces are known as Stone spaces and
Diagram~\ref{eins} becomes
\begin{equation*}
\xymatrix{
  {\ \Stone} \ar@/^/[r]^{P} & {\BA \ } \ar@/^/[l]^{S} .
   }
\end{equation*}
Both diagrams are dual equivalences, so from an abstract point of view
they share exactly the same (or rather dual) properties. But the
categories on the right-hand side are categories of algebras that come
with an equational logic. Lifting such a basic duality to a duality of
coalgebras and modal algebras, as indicated in Diagram~\ref{drei},
will provide modal logics for coalgebras.

\subsection{Notes}

{%

  \noindent References to Stone duality will be given in the next
  section. The standard reference for systems and coalgebras is
  Rutten~\cite{rutten:uc-j}. The duality of algebras/coalgebras and
  induction/coinduction are explained in detail in the tutorial by
  Jacobs and Rutten~\cite{jaco-rutt:tutorial}. Further introductions
  are provided by the course notes of Gumm~\cite{gumm:uc},
  Pattinson~\cite{pattinson:nasslli}, and Kurz~\cite{kurz:esslli01}
  and the forthcoming book by Jacobs~\cite{jacobs:intro-coalg}.

\medskip\noindent\textbf{Coalgebras } 
Motivated by Milner's CCS (4 in Table~\ref{table:coalgebras}), Aczel
\cite{aczel:nwfs} introduced the idea of coalgebras for a functor $T$
as a generalisation of transition systems.  He also made three crucial
observations: (1) coalgebras come with a canonical notion of
bisimilarity; (2) this notion generalises the notion from computer
science and modal logic; (3) any `domain equation' $X\cong TX$ has a
canonical solution (in sets or classes), namely the final coalgebra,
which is fully abstract wrt behavioural equivalence.

\medskip\noindent This idea of a type of dynamic systems being
represented by a functor $T$ and an individual system being an
$T$-coalgebra, led Rutten \cite{rutten:uc-j} to the theory of
universal coalgebra which, parameterised by $T$, applies in a
\emph{uniform} way to a large class of different types of systems.  In
particular, final semantics and the associated proof %
principle of coinduction (which are dual to initial algebra semantics
and induction) find their natural place here.

\medskip\noindent The following references provide details on the
examples in Table~\ref{table:coalgebras}. Stream coalgebras have been
studied by Rutten in a number of papers, see e.g. \cite{rutten:tcs05}.
For the example of deterministic automata as coalgebras see
Rutten~\cite{rutten:automata}.  Probabilistic transition system as
coalgebras go back to Rutten and de
Vink~\cite{vink-rutt:prob-bisim-j}. Coalgebras for the double
contravariant powerset functor are investigated in Kupke and
Hansen~\cite{hans-kupk:cmcs04}.

\medskip\noindent 
The idea of systems as coalgebras and the paradigm of final
semantics---together with its associated
principles of coinduction%
---has been applied to such different topics as, for example, automata
theory \cite{rutten:automata}, combinatorics \cite{rutten:counting},
control theory \cite{kome-schu:dec-sup-con}, denotational semantics of
$\pi$-calculus \cite{FMS96,Sta96}, process calculi and
GSOS-formats
\cite{turi-plot:os,bartels:diss,%
  klin:diss}, probabilistic transition
systems~\cite{bsv:cmcs03-j,cirs-patt:concur04,moss-vigl:cmcs04-j},
component-based software development
\cite{arba-rutt:comp-conn,barbosa:components}, and the solution of
recursive program schemes~\cite{mili-moss:rec}.  Modelling classes in
object-oriented programming as coalgebras \cite{reichel:oo,jacobs:oo}
led to new verification tools (LOOP-Tool \cite{berg-jaco:loop}, CCSL
\cite{rtj:ccsl}, \textsc{CoCasl} \cite{mrrs:cocasl}) which also
incorporate reasoning with modal logics based on the research on
coalgebras and modal logic described below.

\medskip\noindent\textbf{Coalgebras and Modal Logic } For background
on modal logic the reader is referred to Blackburn, de Rijke,
Venema~\cite{brv:ml} (Thm~2.68 shows that modal logic is the
bisimulation invariant fragment of first-order logic, Chapter~5 is on
modal algebras, Thm~4.49 gives an example of an incomplete modal
logic). Further material can be found in Venema~\cite{venema:ac}.
Modal algebras as algebras for a functor and their duality to
coalgebras for a functor was first presented in
Abramsky~\cite{abramsky:bctcs88}.

\medskip\noindent Research into coalgebras and modal logic started
with Moss \cite{moss:cl}.  The logic of \cite{moss:cl} is
uniform\footnote{The restrictions are that $T$ is on $\Set$ and has to
  preserve weak pullbacks.}  in the functor $T$, but it does not
provide the linguistic means to decompose the structure of $T$ which
is needed to allow for a flexible specification language. To address
this issue, \cite{kurz:cmcs98-j,roessiger:ml98-j} (independently)
proposed to restrict attention to specific classes of functors and
presented a suitable, but ad hoc, modal logic.  This work was
generalised by Jacobs \cite{jacobs:many-sorted}.  Pattinson showed
that these languages with their ad hoc modalities arise from modal
operators given by predicate liftings.  He gives conditions under
which logics given by predicate liftings are sound and complete
\cite{pattinson:cml-j} and expressive \cite{pattinson:expressive-j}.
Schr{\"o}der~\cite{schroeder:fossacs05} and Klin~\cite{klin:calco05}
show that for any finitary functor $T$ on $\Set$ there is a modal
logic given by predicate liftings that characterises bisimilarity.

\medskip\noindent 
From a semantical point of view, modal logic can be considered as dual
to equational logic \cite{kurz:aiml98,kurz:diss}.
\cite{kurz-rosi:cmcs02-j} goes further and shows that coalgebras can
be specified---in the same (or dual) way as algebras---by operations
and equations; moreover, the dual of the algebraic operations turn out
to be bisimilarity preserving predicate transformers, that is, modal
formulae. The results following from this approach work for all
functors but the logics need to be strong enough to express all
possible behaviours.  This needs, in general, infinite conjunctions in
the logics. To study finitary logics, Jacobs \cite{jacobs:many-sorted}
covers some ground towards a duality for coalgebras/generalised BAOs
and
Goldblatt \cite{goldblatt:obs-ultraproducts}
develops a notion of ultrapower for coalgebras. Both approaches are
restricted again to specific classes of functors. In this paper we
argue that, based on Stone duality, it is possible to develop a
uniform account.

}%

\section{Stone Duality}\label{sec:sd}

We will treat Stone duality for Boolean algebras as an illustrative
example and then remark on how it generalises to other cases.

\subsection{The Representation Theorem for Boolean Algebras}
\label{sec:sd:rep-thm}

\medskip The axioms of a Boolean algebra relating $\bot, \neg,
\wedge,\vee$ are the abstract essence of the set-theoretic operations
of empty set, complement, intersection and union.  But how can one show
that the axioms of Boolean algebra are indeed complete? We have to
exhibit, for each non-derivable equation, an algebra of subsets
violating that equation. 

\medskip\noindent Suppose $\phi=\psi$ is not derivable from the axioms
of Boolean algebra. By completeness of equational logic, there is a
Boolean algebra $\Asf$ such that $\Asf\notmodels\phi=\psi$. To
conclude that there is a Boolean algebra of subsets (a field of sets)
that refutes $\phi=\psi$ it is enough to find a set $S\Asf$ and an
injective Boolean algebra morphism
\[\Asf \to PS\Asf 
\]
where $P$ denotes here the operation mapping a set to the Boolean
algebra of its subsets. Indeed, if $\Asf\notmodels\phi=\psi$, then by
injectivity $PS\Asf\notmodels\phi=\psi$, yielding a counterexample for
$\phi=\psi$ in an algebra where all the Boolean operations are
interpreted by their set-theoretic counterparts.

\medskip\noindent How does one get the points of the space $S\Asf$? 
Similarly to defining real numbers as certain collections of
intervals, a point will be a certain collection of elements of $\Asf$,
or, equivalently, a function $A\to 2$. Which of these functions should
be points? Observing that $2$ is not only a set, but also a Boolean
algebra $\bbtwo$, we define
$S \Asf = \BA(\Asf,\bbtwo)$
where the notation $\BA(\Asf,\Bsf)$ denotes the set of Boolean algebra
morphisms $\Asf\to\Bsf$.  Detailing the definition of an algebra
morphism, it is straightforward to verify that the requirement that
$p:\Asf\to 2$ be an algebra morphism says that $p$ is a maximal and
consistent collection of elements of $A$.  With the canonical map
$\Asf\to PS\Asf$, we can now state Stone's representation theorem
for Boolean algebras.
Note that with the definition below, the statement that a point $p$
satisfies the predicate $a$ expresses itself as $p\in\hat a$.
\begin{theorem}\label{thm:rep-BA}
The map
 \begin{align}
   \hat{(\cdot)}: \Asf & \too P S \Asf\\
   a & \ \mapsto \ \hat a = \{ p \in S \Asf \mid p(a)=1\} \label{equ:def-hat}
 \end{align}
 is an injective Boolean algebra morphism.
\end{theorem}

\subsection{Stone Duality for Boolean Algebras}
\label{sec:sd:duality}

The representation theorem works by associating a space to an algebra
(via $S$) and then, vice versa, an algebra to a space (via $P$).
What precisely are the spaces that correspond to algebras?

\medskip\noindent In a first instance, we can say that a space
$(X,\Asf)$ consists of a set $X$ and a Boolean algebra of subsets
$\Asf\subseteq 2^X$ such that (1) any two different points in $X$ are
separated by elements of $\Asf$ and (2) $(X,\Asf)$ is compact, that
is, every collection $\ccal$ of elements of $\Asf$ with the finite
intersection property \footnote{$\ccal$ has the finite-intersection
  property if all finite subset of $\ccal$ have non-empty
  intersection.} has non-empty intersection.  The two properties
capture that the points of the space are determined by the algebra in
the following sense. (1) says that there are not more points than can
be separated by predicates and (2) that there are enough points
to realise every consistent collection of predicates from $\Asf$.

\medskip\noindent Further, one notices that a space $(X,\Asf)$ can be
considered as the topological space $(X,\ocal X)$ with $\ocal X$ being
the topology generated by $\Asf$, that is, the closure of $\Asf$ under
arbitrary unions. One recovers the Boolean algebra $\Asf$ from $\ocal
X$ as the collection of all compact opens. Since in a compact
Hausdorff space a subset is compact iff it is closed, one can replace compact open by clopen (which is brief for closed and open). To summarise:

\begin{definition}
  A Stone space is a topological space that (1) is $T_0$, (2)
  compact, and (3) the clopens are a basis for the topology.
\end{definition}

\noindent Stone spaces with continuous maps form the category $\Stone$.
From the representation theorem and the definition of $\Stone$ we
obtain two operations $S$ and $P$
\[
\xymatrix{
\Stone \ar@/^/[r]^{P}& \BA\ar@/^/[l]^{S}
}
\]
where $S\Asf$ is the topological space with points $\BA(\Asf,\bbtwo)$
and the topology generated by $\{\hat a\mid a\in \Asf\}$ as in
(\ref{equ:def-hat}); $PX=\Stone(X,2)$ is now the Boolean algebra of
clopens (instead of the full powerset).  We speak of a duality here
because both operations are functors that act on morphism by reversing
the arrows, namely, mapping a morphism $f$ to inverse image $f\inv$.
Moreover, $\Stone$ and $\BA$ are dually equivalent, that is we have
isomorphisms
\begin{align}
\Asf & \iso PS\Asf \label{A} \\
X & \iso SPX \label{X}
\end{align}
(\ref{X}) is injective because $X$ is $T_0$ and surjective
because $X$ is compact. (\ref{A}) is surjective by construction and
injective by the Representation Theorem~\ref{thm:rep-BA}.  To
summarise:

\begin{theorem}
  The categories $\BA$ and $\Stone$ are dually equivalent.
\end{theorem}

\noindent From our presentation, one could get the impression that
topologies come in here accidentally and the logical content of the
duality is completely contained in the representation theorem. I would
reply the following.  First, Stone spaces arise here from logical
considerations but they are of independent interest. A well-known
example is the Cantor middle-third space. In fact, all complete
ultrametric spaces are Stone spaces. Second, the dual equivalence is
nice to have; for example, we then have that the dual of an initial
algebra is the final coalgebra; this will be used in the next section
to show that that modal logics characterise bisimilarity.  Third,
topologies often have an interesting computational perspective arising
from the idea that observable properties are closed under arbitrary
unions but not intersections
\cite{smyth:handbook,vickers:tvl,escardo:st}.  Finally, the
topological perspective suggests and unifies many generalisations,
some of which we briefly review now.

\subsection{A Sketch of the  General Picture}
\label{sec:sd:general}

The variations of Stone duality relevant for the present purposes fit
the following picture. We start with a class $\acal$ of distributive
lattices and $\xcal$ of topological spaces (assumed to be $T_0$).
Think of algebras $A\in\acal$ as propositional theories and of spaces
$X\in\xcal$ as models of propositional theories with the opens (or
compact opens for finitary logics) interpreting the propositions.
There is an operation $P:\xcal\to\acal$, mapping a space to its
algebra of predicates. And an operation $S:\acal\to\xcal$ mapping an
algebra to its `canonical model'.
\begin{equation}\label{equ:basic}
\xymatrix{ 
  {\ \xcal} \ar@/^/[r]^{P} & {\acal \ } \ar@/^/[l]^{S}. 
   } 
\end{equation}
Moreover, in the examples of the table below, $PX=\xcal(X,2)$ and
$SA=\acal(A,2)$ where 2 denotes the appropriate two-element
topological space or two-element algebra.  We speak of a duality,
since $P$ and $S$ work contravariantly on morphisms, mapping a
morphism (that is, algebra morphism or continuous map) $f$ to $f\inv$.
Moreover, there are morphisms
\[ \rho_\Asf: \Asf\to PS\Asf \quad\quad\quad\quad \sigma_\Xsf: \Xsf\to SP\Xsf \]
and (\ref{equ:basic}) is a dual equivalence if they are bijective.
Logically, this means the following. As we have explained in
Section~\ref{sec:sd:rep-thm}, $\rho_\Asf$ injective means completeness
(and it will, in general, be surjective by definition of $S$ and $P$).
$\sigma_\Xsf$ is injective means, together with $\Xsf$ being $T_0$,
that the logic is expressive in the sense that different points are
separated by some predicate. If $\sigma_\Xsf$ is not surjective, then
$SPX$ has points not available in $X$; thus the logic is not strong
enough to make these additional points inconsistent.

\medskip\noindent We conclude with a table of some relevant examples.

\renewcommand{\arraystretch}{1.4}
\begin{center}
\begin{tabular}{|l|l|l|l|}
  \hline
  $\xcal$ & $\acal$ & spaces/algebras & propositional logic \\
  \hline
  \hline
  $\Set$ & $\CABA$ & sets/complete atomic Boolean algebras & infinitary classical \\
  \hline
 $\Stone$ & $\BA$ & Stone spaces/Boolean algebras & classical \\
  \hline
 $\Spec$ & $\DL$ & spectral spaces/bounded distributive lattices & negation free \\
  \hline
 $\Poset$ & $\CDL$ & posets/complete distributive lattices  & infinitary negation free \\
  \hline
 $\Sob$ & $\Frm$ & sober spaces/frames   & geometric \\
  \hline
\end{tabular}
\end{center}

\noindent In the two last examples, $\rho_\Asf$ is injective for free
algebras $\Asf$ but not for all algebras. Logically, this corresponds
to having completeness but not strong completeness. This also happens
for propositional logic with countable conjunctions.

\subsection{Notes}

{%

  Three introductory textbooks on Stone duality are
  Vickers~\cite{vickers:tvl}, Davey and Priestley \cite{dave-prie:lo},
  Brink and Rewitzky~\cite{brin-rewi:ps}.

  \medskip\noindent Stone duality was introduced by Stone
  \cite{stone:ba, stone:dl}. The main reference for Stone duality is
  Johnstone's book on Stone Spaces \cite{johnstone:stone-spaces} which
  also provides detailed historical information. The handbook article
  \cite{abra-jung:dt} covers the topic from the point of view of
  domain theory. Both texts also provide many more examples of Stone
  dualities. Topological dualities beyond sober spaces, e.g., for
  completely distributive lattices and posets, are treated by
  Bonsangue et al~\cite{bjk:tzero,bonsangue:diss}. The representation
  theorem for propositional logic with countable conjunctions can be
  found in Karp~\cite{karp:infinite}. For applications of complete
  ultrametric spaces to control flow semantics see de Bakker and de
  Vink~\cite{bakk-vink:cfs}. 
}

\section{Logics of Coalgebras}\label{sec:lc}

The previous section discussed dual equivalences (\ref{equ:basic})
between categories $\xcal$ of topological spaces and categories
$\acal$ of distributive lattices.
In this section, we extend this picture to $T$-coalgebras. Starting
with a diagram as in (\ref{equ:basic}) and a functor $T$ on $\xcal$,
we dualise $T$ to a functor $L$ on $\acal$. The duality of
$\xcal$/$\acal$ and $T/L$ lifts to a duality of coalgebras and
algebras.
\begin{equation}\label{diag:XA}
\xymatrix{
  {\Coalg(T)} \ar@/^/[r]^{\fca} \ar[d]
  & {\Alg(L)\ } \ar@/^/[l]^{\fac} \ar[d]\\
  {\ \cal X} \ar@(dl,ul)[]^{T} \ar@/^/[r]^{P}
   & {\acal\ } \ar@/^/[l]^{S} \ar@(dr,ur)[]_{L} 
  \\
}
\end{equation}
And in the same way as the duality of $\xcal$ and $\acal$ describes a
logic for $\xcal$, so the duality of $\Coalg(T)$ and $\Alg(L)$
describes a logic for $T$-coalgebras.

\subsection{Abstract Logics: Using the Duality}\label{sec:lc:abstract}

\medskip\noindent Given a duality as in (\ref{equ:basic}) and a
functor $T$ on $\xcal$, then $PTS$ is the dual of $T$ on $\acal$. In
fact, we will need a bit more liberty and say that $L$ is dual to $T$
if $L$ is isomorphic to $PTS$. Or, equivalently, $L$ is
\emph{dual} to $T$ if there is a natural isomorphism
\begin{equation}\label{delta}
\delta_X: LPX\to PTX
\end{equation}
Using $\delta$ we can associate to a $T$-coalgebra $(X,\xi)$ its dual
$L$-algebra
\begin{gather*}
  \fca(X,\xi) \ = \    LPX\stackrel{\delta_X}{\too} PTX \stackrel{P\xi}{\too} PX\end{gather*}
and similarly for $S$.

\medskip\noindent In algebraic logic, logics are described by
operations and equations, and then properties of a logic are studied
by investigating the variety of the algebras for the given operations
and equations. A basic construction is that of the Lindenbaum algebra.
Given a logic $L$, the Lindenbaum algebra $A_L$ is obtained from
quotienting the set of all terms by the smallest congruence derived
from the equations.  Thus, the elements of the Lindenbaum algebra
$A_L$ can be seen as `abstract propositions', or propositions up to
interderivability.  Among all algebras in the variety, the Lindenbaum
algebra is determined by the following property: for any algebra $A$
there is a unique morphism $A_L\to A$, that is, $A_L$ is the initial
algebra.  We turn this into a definition. 

\begin{defn}\label{def:L-logic}
  Denote by $A_L$ the initial $L$-algebra. The elements of $A_L$ are
  called propositions.  The semantics $\sem{\phi}_{(X,\xi)}$ of a
  proposition $\phi$ wrt a coalgebra $(X,\xi)\in\Coalg(T)$ is given by
  the image of $\phi$ under
  \[ A_L\too \tilde P (X,\xi) \]   
  We write $\Coalg(T)\models (\phi=\psi)$ if for all coalgebras
  $(X,\xi)$ the equation $\phi=\psi$ is satisfied in $\tilde P (X,\xi)$.
\end{defn}

\noindent We remark that Theorem~\ref{thm:present-fun} will explain
precisely in what sense the initial $L$-algebra is a Lindenbaum
algebra.

\begin{theorem}%
\label{thm:invariance}
Propositions are invariant under bisimilarity.
\end{theorem}

\begin{proof}
  Recalling the definition of bisimilarity (p.~\pageref{def-bisim}), we
  have to show, given a coalgebra morphism
  $f:(X,\xi)\to(X',\xi')$ 
  and $x\in X$, that
  $x\in\sem{\phi}_{(X,\xi)} \ssi f(x)\in\sem{\phi}_{(X',\xi')}$.
  This follows directly from the fact that the diagram
  \[
  \xymatrix@R=1pc{
    & \fca(X,\xi)\\
    A_L \ar [ur]^{\sem{-}_{(X,\xi)}\ } 
                    \ar [dr]_{\sem{-}_{(X',\xi')}} 
                    & \\
    & \fca(X',\xi')\ar[uu]_{\fca f = Pf = f^{-1}}
  }
  \]
 commutes due to $A_L$ being initial.
\end{proof}

\noindent The essence of completeness wrt to the coalgebraic semantics is: 

\begin{theorem}%
\label{thm:equivalence} 
${\Alg(L)} \models (\phi = \psi) \ \ssi\ {\Coalg(T)}\models (\phi=\psi)$.
\end{theorem}

\begin{proof}
  `$\oif$' (soundness) is immediate from the definitions. `$\si$'
  (completeness) works as in Theorem~\ref{thm:rep-BA}. Suppose
  $A_L\notmodels\phi=\psi$. By injectivity of $A_L\to \tilde P\tilde
  SA_L$ we have $\tilde P\tilde SA_L\notmodels\phi=\psi$. That is, the
  coalgebra $\tilde SA_L$ does not satisfy $\phi=\psi$.
\end{proof}

\noindent We remark that, as apparent from the proof, it is the
representation of the initial (or, more generally, free algebras)
which gives completeness. Since we have a dual equivalence, all
algebras can be represented and we obtain strong completeness
(completeness wrt a set of assumptions).

\begin{theorem}\label{thm:expressive}
  The logic characterises bisimilarity. 
\end{theorem}

\begin{proof}
  Without loss of generality, let us assume that $x,x'$ are two
  different elements of the final coalgebra $\tilde SA_L$. The two
  points can be distinguished by a proposition since $ A_L\to\tilde
  P\tilde SA_L$ is surjective and $\tilde SA_L$ is a $T_0$-space.
\end{proof}

\medskip\noindent To summarise, we have seen how to obtain a logic
that perfectly describes $T$-coalgebras: Just consider as formulae the
elements of the initial $L$-algebra\footnote{If $T:\Set\to\Set$ is
  powerset, then the initial $L$-algebra does not exist for reasons of
  size. But one can still define a class of formulae using the initial
  algebra sequence of $L$. We ignore this slight complication for the
  purposes of exposition.}  where $L$ is the dual of $T$. We called
this logic abstract since it is not explicitly built from modal
operators and axioms. The next subsection explains that modal
operators and axioms are presentations of the functor $L$.

\subsection{Concrete Logics: Presenting Algebras and Functors}
\label{sec:lc:present}

Ultimately, we are interested in relating logical calculi to
transition systems. We have motivated to consider transition systems
as coalgebras and used Stone duality to dualise coalgebras to
algebras. The particular benefit obtained from using Stone duality is
that the algebras thus obtained correspond to logical calculi.  Let us
take a closer look again at the guiding ideas, which have been:

\renewcommand{\arraystretch}{1.4}
\begin{center}
\begin{tabular}[h]{|l|l|}
  \hline
  category of algebras $\acal$ & propositional logic\\
  \hline
  algebra $A$ in $\acal$ & propositional theory \\
  \hline
  functor $L:\acal\to\acal$ & operations and equations for $T$-coalgebras\\
  \hline
  category $\Alg(L)$ & modal logic for $T$-coalgebras\\
  \hline
\end{tabular}
\end{center}

\noindent These correspondences are justified as follows. The
categories $\acal$ obtained from Stone duality can be presented by a
signature $\Sigma$ of operations and equations $E$ in the sense that
$\acal$ is (isomorphic to) the class $\Alg(\Sigma,E)$ of algebras for
the signature $\Sigma$ satisfying $E$. The presentation $\langle
\Sigma,E\rangle$ gives a logical calculus, via equational logic.
An algebra $A\in\acal$ has a presentation $\langle G,R\rangle$ by
generators and relations if $A$ is isomorphic to the quotient of the
free algebra over $G$ by the smallest congruence containing $R$. In
our context, this means that $A$ is the propositional theory given by
variables $G$ and additional axioms $R$, see the example below.

\medskip\noindent\textbf{Presenting functors } We emphasised above the
point of view that a propositional logic is a presentation of a
category of algebras. Similarly, it is a presentation of $L$ that
gives rise to the modal operators and its axioms.

\medskip\noindent\emph{Example. } The functor $L:\BA\to\BA$ for
$\pcal$-coalgebras from (\ref{equ:def-L}) is presented by the unary
operator $\Box$ and the equations ($\ref{K}$) in the following sense.
For each $A\in\BA$, the algebra $LA$ is presented by generators
$\{\Box a\mid a\in A\}$ and by relations
$\{(\Box\top,\top)\}\cup\{(\Box(a\et b),\Box a\et \Box b)\mid a,b\in
A\}$.

\medskip\noindent It is not a coincidence that the equations in this
example are of a special format: Roughly speaking, they do not allow
nesting of modal operators. Such terms are called terms of rank~1:

\begin{definition}
  Assume $\acal\iso\Alg(\Sigma,E)$ and a signature $\Sigma'$ (with
  operation symbols disjoint from $\Sigma$). A term in
  $\Sigma+\Sigma'$ is of rank~1 (wrt $\Sigma'$) if it is of the form
  $t(\Box_i(s_{ij}))$ where $t$ is an $n$-ary term in $\Sigma$ and the
  $\Box_i$, $0\le i < n$, are $m_i$-ary operations in $\Sigma'$ and
  the $s_{ij}$, $0\le j<m_i$ are terms in $\Sigma$. An equation $t=s$
  is of rank~1 if both terms are.
\end{definition}

\noindent In our example, the equations~(\ref{K}) are of rank~1. In
particular: $\top$ is a term of rank~1, because $\top$ is a 0-ary term
in the signature $\Sigma$ of Boolean algebras; $\Box(a\et b)$ is a
term of the form $t(\Box(s))$ where $t$ is a variable and $s$ is $a\et
b$. Terms like $\Box a \imp a$ and $\Box a \imp\Box\Box a$ are not of
rank~1. We can now define what it means to present a functor by
operations and equations.

\begin{definition}
  Assume $\acal\iso\Alg(\Sigma_\acal,E_\acal)$, a signature $\Sigma_L$
  and a set of equations $E_L$ that are of rank~1 (wrt $\Sigma_L$).
  $\langle \Sigma_L, E_L\rangle$ is a presentation of
  $L:\acal\to\acal$ if the algebras $LA$ are presented by $\langle
  G_A, R_A\rangle$ where $G_A=\{\sigma(a_i)\mid \sigma\in\Sigma_L,
  a_i\in A\}$ and $R_A$ consists of all substitution instances of
  equations in $E_\acal\cup E_L$ obtained by replacing variables with
  elements from $A$.
\end{definition}

\medskip\noindent Generalising the example above, it now follows that
logics given by predicate liftings correspond to functors
$L:\BA\to\BA$. Indeed, if $\Sigma$ is a collection of predicate
liftings (with arities possibly $>1$), then $\langle
\Sigma,\emptyset\rangle$ presents some functor $L$. Moreover, it is
not hard to see that the two semantics of the modal operators given by
(\ref{sem-box}) and Definition~\ref{def:L-logic} coincide. This also
means that, conversely, any presentation of a functor corresponds to a
collection of predicate liftings (given by $\Sigma_L$) plus some
additional axioms. 

\medskip\noindent The next theorem links the abstract logics from the
previous section with concrete logical calculi. In particular, it
shows that the Lindenbaum algebra of the logic given by operations
$\Sigma_\acal+\Sigma_L$ and equations $E_\acal+E_L$ is the initial
$L$-algebra. The proof that every $L$-algebra satisfies the equations
$E_L$ requires the restriction to rank~1.

\begin{theorem}\label{thm:present-fun}
  Assume $\acal\iso\Alg(\Sigma_\acal,E_\acal)$ and $L:\acal\to\acal$.
  If $L$ has a presentation $\langle \Sigma_L, E_L\rangle$ then
  $\Alg(L)\iso\Alg(\Sigma_\acal+\Sigma_L, E_\acal+E_L)$.
\end{theorem}

\medskip\noindent The theorem can be read in two directions. First,
starting with $T$, we find a presentation for $L$ and obtain
completeness results for modal calculi. Of course, finding a good such
presentation for a functor is usually not straightforward. It is
therefore of interest to know whether arbitrary functors $L$ do have a
presentation. This question has recently received a positive answer
for finitary functors on Boolean algebras and sifted colimits
preserving functors on arbitrary varieties.
I expect that these results can be extended to show
that modal calculi exist for all functors $T$ on $\Set$ and related
categories.

\medskip\noindent Second, one can take a logical calculus and study
the corresponding presented functor. For example, the infinitary
version of the modal calculus $\textbf{K}$ presents the dual
$L:\CABA\to\CABA$ of $\pcal:\Set\to\Set$. We obtain the theorem
mentioned in the introduction that infinitary modal logic
characterises bisimilarity.  Moreover, we also get a strong
completeness result for the infinitary version of the modal calculus
$\textbf{K}$. For $\textbf{K}$ itself we obtain the corresponding
results for the powerspace (also known as Plotkin power domain or
hyperspace) on $\Stone$. Similarly, any modal logic of rank~1 is
expressive and strongly complete for some functor of Stone spaces
(also this may not be intended semantics). This methodology can be
applied to all functors in Table~\ref{table:coalgebras} as
presentations of their duals are known, with the possible exception of
the functors for probabilistic transition systems which deserve some
further attention.

\subsection{Notes}

{%

  \medskip\noindent The application of Stone duality to modal logic
  goes back to J\'onsson and
  Tarski~\cite{jons-tars:bao1,jons-tars:bao2} and then
  Goldblatt~\cite{goldblatt:meta-76}. The idea of relating type
  constructors on algebras (see the $L$ above) and topological spaces
  (see the $T$ above) is from Abramsky's Domain Theory in Logical Form
  \cite{abramsky:dtlf,abramsky:deb}. Compared to
  \cite{abramsky:dtlf,abramsky:deb}, the models we are interested in
  are not only solutions to recursive domain equations (final
  coalgebras) but any coalgebras; moreover, their base category need
  not be a domain but can be a more general topological space.
  Compared to \cite{goldblatt:meta-76}, we use the duality of algebras
  and coalgebras to lift the Stone duality from Boolean logic to modal
  logic. Our functors $L$ (or their presentations) are closely related
  to C\^{\i}rstea's language and proof system constructors
  \cite{cirstea:cmcs03}.

  \medskip\noindent \cite{kkv:cmcs03-j} studies coalgebras over Stone
  spaces to show that they capture the descriptive general frames from
  modal logic and to present a different view on Jacobs many-sorted
  coalgebraic modal logic~\cite{jacobs:many-sorted};
  \cite{palmigiano:cmcs03-j} applies this approach to give a
  coalgebraic analysis of positive modal logic;
  \cite{bons-kurz:fossacs05} proposes to study logics for coalgebras
  via the dual functor and shows that powerspace can be treated in a
  uniform way for different categories of topological spaces;
  \cite{bons-kurz:fossacs06} introduces the notion of a functor
  presented by operations and equations; \cite{kkp:cmcs04} shows that
  logics given by predicate liftings can be described by functors
  $L:\BA\to\BA$; \cite{kkp:calco05} studies the relationship between
  Stone-coalgebras and Set-coalgebras.

}

\section{Outlook}\label{sec:outlook}

The aim of this exposition was to give a principled explanation of 
coalgebras and their logics. It cannot be denied that it took us some
work in Section~\ref{sec:sd} to set up the necessary machinery. On the
other hand, we got paid back with short and easy proofs of
Theorems~\ref{thm:equivalence} and \ref{thm:expressive}. Notice that
these proofs of completeness and expressiveness do not involve any
syntax. The interface between syntax and semantics, so to speak, is
provided by the notion of a presentation of a functor. This provides
an interesting way to reason about different modal logics in a uniform
and syntax independent way.

\medskip\noindent One of the benefits of setting up the theory of
coalgebras and their logics in a way uniform in the functor is
compositionality. For example, given presentations for $L_1$ and
$L_2$, one obtains a presentation of the composition $L_1 L_2$. This
allows us to not only build new types of coalgebras from old ones, but
to do the same for their associated logics (as done already in
Abramsky~\cite{abramsky:dtlf}). The power of this approach is
exemplified by \cite{bons-kurz:pi-log} which derives a logic for
$\pi$-calculus: Using known results and compositionality, a
presentation for the functor of $\pi$-calculus is not difficult to
find and we can then apply the general results.

\medskip\noindent Let us conclude with some further topics.

\medskip\noindent\textbf{The modal logic of a functor } Our original
question has been the following. If universal coalgebra is a general
theory of systems as proposed by Rutten~\cite{rutten:uc-j}, then what
are the logics for coalgebras? More specifically, can the theory of
logics for $T$-coalgebras be developed uniformly in the functor $T$?
The insight alone that, semantically, modal logic is dual to
equational logic \cite{kurz:diss} does not give a handle on relating
coalgebras and their modal calculi. As shown here, this is where Stone
duality comes in. The solution to the original problem of associating
a logic to a functor $T$ now looks in close reach: It will be shown
that, under appropriate conditions, the dual of a functor $T$ has a
presentation, which then provides a strongly complete modal logic
characterising bisimilarity.  This should also allow to generalise
Moss's original work \cite{moss:cl} and provide his logic with a complete
calculus.

\medskip\noindent\textbf{Relating different Stone dualities }
Topology-based models arise either, as in this article, to capture the
expressivity of logics weaker than infinitary classical logic, or in
situations, as in domain theory, where a natural notion of observable
predicate is given. In both cases, it would be interesting to be able
to treat the topology as a parameter. This would allow us to compare
similar models based on different categories of spaces and to study
logics which involve two different Stone dualities, e.g., the ones for
$\BA$ and $\Set$. Ongoing work is based on the idea to consider both
dualities as arising from different completions of one and the same
simpler duality.

\medskip\noindent\textbf{Logics with name binding } The work on the
logic of $\pi$-calculus \cite{bons-kurz:pi-log} suggests that also
other logics with name binding and quantifiers can be usefully treated
in the presented framework. This needs still to be worked out.

\medskip\noindent\textbf{Coalgebraic modal model theory } In order to
better appreciate the relationship between modal logic and coalgebras,
it would be good to understand in how far known results in modal logic
can be extended to coalgebras. Some work in this direction has been
done on the J\'onsson-Tarski-theorem and ultrafilter extensions
\cite{kkp:calco05}.  There are also new questions brought to modal
logic from coalgebra, for example, how to best deal with infinite
parameters $C$ in Table~\ref{table:coalgebras}, see Friggens and
Goldblatt~\cite{frig-gold:polynomial}.

\medskip\noindent\textbf{Going beyond rank~1 } The original motivation
in using Stone dualities was to understand logics of coalgebras for a
functor $T$. We have seen that a logic for $T$ only needs axioms of
rank~1.  From this point of view, rank~1 is no restriction.  And, of
course, we can deal with axioms of rank $>1$ in a trivial way: axioms
of rank~1 determine a functor $T$ and hence a category $\Coalg(T)$,
whereas the other axioms specify a subcategory of $\Coalg(T)$. So the
question really is whether axioms not of rank~1 can be treated in a
\emph{uniform coalgebraic} way.

\medskip\noindent\textbf{Fixed-point logic } It is straightforward to
extend a basic logic derived from $T$ by fixed-points as in
$\mu$-calculus. But it is not clear at all whether a Stone duality
based approach can help in better understanding fixed-point logics.
Venema~\cite{venema:coalg-aut} and Kupke and
Venema~\cite{kupk-vene:lics05} introduce the notion of coalgebraic
fixed point logic and show that $\mu$-calculus interpreted over
different data structures such as words and trees can be treated
uniformly in a coalgebraic framework.

\appendix\section{Some Notions of Category Theory}

A \emph{category} $\ccal$ consists of a class of objects and has, for
any two objects $A,B$, a set $\ccal(A,B)$ of arrows (or morphisms)
from $A$ to $B$. Furthermore, arrows $f:A\to B, g:B\to C$ have a
composition $g\circ f$ and for each object $A$ there is an identity
arrow $\id_A$.  Examples: The category $\Set$ with sets as objects and
functions as arrows; $\BA$ with Boolean algebras and their
homomorphisms; further, $\Coalg(T)$ and $\Alg(L)$.

\medskip\noindent An \emph{isomorphism} is an arrow $f:A\to B$ for
which there is a $g:B\to A$ with $f\circ g=\id_B, g\circ f=\id_A$.

\medskip\noindent A \emph{covariant functor} between two categories
$F:\ccal\to\dcal$ maps objects to objects and arrows $f:A\to B$ to
$Ff:FA\to FB$, preserving identities and composition. Examples: the
functors $T$ and $L$. 

\medskip\noindent For each category $\ccal$ we have the \emph{dual
  category} $\ccal\op$ obtained from reversing the arrows. Example:
Each functor $F:\ccal\to\ccal$ gives rise to a functor
$F\op:\ccal\op\to\ccal\op$; the duality of algebras and coalgebras can
now be stated as $\Alg(F\op)=\Coalg(F)\op$.

\medskip\noindent A \emph{contravariant functor} $F:\ccal\to\dcal$ is
a covariant functor $\ccal\op\to\dcal$ or, equivalently,
$\ccal\to\dcal\op$, that is, it reverses the direction of the arrows.
Example: $2^-:\Set\to\BA$ maps $f:X\to Y$ to $2^f=f\inv:2^Y\to 2^X$.

\medskip\noindent Given functors $F,G:\ccal\to\dcal$, a \emph{natural
  transformation} $\tau:F\to G$ consists of maps $\tau_A:FA\to GA$,
$A$ in $\ccal$, such that for all $f:A\to A'$ we have
$Gf\circ\tau_A=\tau_{A'}\circ f$.  Example: the predicate
liftings~(\ref{pred-lift}). 

\medskip\noindent Given two functors $F:\ccal\to\dcal$ and
$G:\dcal\to\ccal$, we say that $\ccal$ and $\dcal$ are
\emph{equivalent} if there are natural isomorphisms $\tau_C:C\to
GFC$ and $\sigma_D:D\to FGD$.  $\ccal$ and $\dcal$ are \emph{dually
  equivalent} if $\ccal\op$ and $\dcal$ are equivalent.

\end{document}